\newif\ifSV
\SVfalse
\newif\iffullpage
\fullpagefalse
\ifSV
\documentclass[nospthms]{svjour3}
\else
\documentclass[publicationstatus]{eptcs}
\providecommand{\event}{SR 2015}
\fi
\newif\ifLONG
\LONGtrue
\usepackage{a4}
\usepackage[all]{xy}
\iffullpage \usepackage{fullpage}\fi
\usepackage{qsymbols}
\usepackage{marvosym}
\usepackage{fancybox}
\usepackage{url}
\usepackage{color}
\usepackage{amsmath,wasysym}
\usepackage{graphicx}

\newcommand{\All}{\textsf{A}} 
\newcommand{\Bel}{\textsf{B}}
\newcommand{\Al}{\mbox{\footnotesize \textsf{A}}}
\newcommand{\Be}{\mbox{\footnotesize \textsf{B}}}

\newcommand{\prof}[1]{ \langle\!\langle #1\rangle\!\rangle}
\newcommand{\conv}[1]{\;\downarrow#1}
\newcommand{\diverg}[1]{\;\uparrow#1}
\newcommand{\Conv}[1]{\Box\!\downarrow#1}
\newcommand{\Ratf}{\ensuremath{\mathsf{Rat}_{f}}}
\newcommand{\Ratoo}{\ensuremath{\mathsf{Rat}_{\infty}}}

\newcommand{\nat}{\ensuremath{\mathbb{N}}}
\newcommand{\real}{\ensuremath{\mathbb{R}}}

\newcommand{\nodA}{*++[o][F]{\Al}}

\newcommand{\nodB}{*++[o][F]{\Be}}
\newcommand{\fl}[1]{\ar@/^/[#1]^r \ar@/^/[d]^{d}}
\newcommand{\flr}[1]{\ar@[blue]@2@/^/[#1]^{\color{blue} 2} \ar@/^/[d]^{1}}
\newcommand{\fld}[1]{\ar@/^/[#1]^2 \ar@[blue]@2@/^/[d]^{\color{blue} 1}}

\newcommand{\infpede}{\ensuremath{\infty}pede}
\newcommand{\ompede}{\ensuremath{`w}pede}

\newcommand{\titlerunning}{The risk of divergence}
\newcommand{\authorrunning}{P. Lescanne}

\newenvironment{proof}[1]{\begin{quotation}\noindent\textsf{Proof:} #1}{\(\Box\)\end{quotation}}
\newtheorem{theorem}{Theorem}
\newtheorem{proposition}[theorem]{Proposition}

\newtheorem{definition}[theorem]{Definition}

\begin{document}
\ifSV
\title{The risk of divergence}
\author{Pierre Lescanne}
\institute{University of Lyon, \'Ecole normale sup\'erieure de Lyon, CNRS (LIP), \\ 46 all\'ee
d'Italie, 69364 Lyon, France}
\else
\title{The risk of divergence}
\author{Pierre Lescanne\\
University of Lyon, \'Ecole normale sup\'erieure de Lyon, CNRS (LIP), \\ 46 all\'ee
d'Italie, 69364 Lyon, France}
\fi

\maketitle

\begin{abstract}
\ifSV\else  \medskip\hrule\fi

\medskip

We present infinite extensive strategy profiles with perfect information and we show
that replacing finite by infinite changes the notions and the reasoning tools.  The
presentation uses a formalism recently developed by logicians and computer science
theoreticians, called coinduction.  This builds a bridge between economic game theory
and the most recent advance in theoretical computer science and logic.  The key
result is that rational agents may have strategy leading to \emph{divergence}.

\newcommand{\kwds}{divergence, decision, infinite game, sequential game,
  coinduction.}

\ifSV\end{abstract}
\keywords \\kwds
\else \medskip

\end{abstract}
\hrule\fi

\section{Introduction}
\label{sec:intro}

Strategies are well described in the framework of sequential games, aka.  games in
extensive forms with perfect information.   In this paper, we describe rational strategies
leading to divergence.\footnote{In this paper we use ``divergence'' instead of
  ``escalation'' since it is somewhat dual convergence a concept which plays a key
  role in what follows.}  Indeed
divergence understands that the games, the strategies and the strategy profiles are
infinite.  We present the notion of infinite strategy profiles
together with the logical framework to reason on those objects, namely coinduction.

\section{Decisions in Finite Strategy Profiles}
\label{sec:fin_Strat_Prof}

To present strategy reasoning, we use one of the most popular framework, namely
\emph{extensive games with perfect information} (\cite{osborne04a} Chapter 5 or \cite{shaun04:_game_theor}) and we
adopt its terminology.  In particular we call \emph{strategy profile} an
\emph{organized} set of strategies, merging the decisions of the agents.  This
organization mimics this of the game and has the same structure as the game
itself.  They form the set\footnote{To be correct, we should say the \emph{``they
    form the coalgebra''}.}  \textsf{StratProf}.  By \emph{``organized''}, we mean
that the strategic decisions are associated with the nodes of a tree which correspond
to positions where agents have to take decisions.  In our approach strategy profiles
are first class citizens and games are byproduct. In other words, strategy profiles
are defined first and extensive games are no more than strategy profiles where all
the decisions have been erased.  Therefore we will only speak about strategy
profiles, keeping in mind the underlying extensive game, but without giving them a
formal definition\footnote{A direct definition of games is possible, but is not necessary in
  this paper.}.  For simplicity and without loss of generality, we consider only
dyadic strategy profiles (i.e.; double choice strategy profiles) , that are strategy profiles with only two choices at each
position.  Indeed it is easy to figure out how multiple choice extensive strategy
profiles can be represented by double choice extensive strategy profiles.  We let the
reader imagine such an embedding.  Therefore, we consider a set of choices:
$\mathsf{Choice} = \{1,2\}$.

Along the paper, our examples need only a set of two agents: $\mathsf{Agent} =
\{\All, \Bel\}$.  In this paper we use coinduction and corecursion as basic tools for
reasoning correctly about and defining properly infinite objects.  Readers who want
to know more about those concepts are advised to read introductory
papers~\cite{jacobs12:_introd_coalg,DBLP:journals/tcs/Rutten00}, while specific
applications to infinite strategy profiles and games are introduced in
\cite{lescanne14:_intel}.

\begin{definition}
  A \emph{finite strategy profile} is defined \emph{by induction} as follows:
  \begin{itemize}
  \item either given a \emph{utility assignment} $u$ (i.e., a function $u:
    \mathsf{Agent} "->" \real$) $\prof{u}$ is a \emph{finite strategy profile}, which
    corresponds to an ending position.
  \item or given an \emph{agent} $a$, a \emph{choice} $c$ and two \emph{finite strategy profiles} $s_1$ and
    $s_2$, $\prof{a, c, s_1, s_2}$ is a \emph{finite strategy profile}.
  \end{itemize}
\end{definition}

For instance, a strategy profile can be drawn easily with the convention that $1$ is
represented by going down and $2$ is represented by going right. The chosen
transition is represented by a double arrow
$\xymatrix@C=10pt{\ar@[blue]@{=>}@/^.3pc/[rr]&&}$.  The other transition is
represented by a simple arrow $\xymatrix@C=10pt{\ar@/^.3pc/[rr]&&}$.  For instance
\begin{displaymath}
           \xymatrix@C=10pt{
          &\ar@{.>}[r]& *++[o][F]{\Al} \ar@[blue]@{=>}@/^1pc/[rr]^2 \ar@/^/[d]^1
          &&*++[o][F]{\Be} \ar@/^1pc/[rr]^2 \ar@[blue]@{=>}@/^/[d]^1  && 0,5\\
          &&1,0.5&&2,1
        }
\end{displaymath}
is a graphic representation of the strategy profile \[s_{`a} = \prof{\All, 2,
  \prof{\All"|->"1,\Bel"|->"0.5},\prof{\Bel,1,\prof{\All"|->"2,\Bel"|->"1},\prof{\All"|->"0,\Bel"|->"5}}}.\]

From a finite strategy profile, say $s$, we can define a \emph{utility
  assignment}, which we write $\widehat{s}$ and which we define as
follows:
\begin{itemize}
\item $\widehat{\prof{u}} = u$
\item $\widehat{\prof{a,c,s_1,s_2}} = \mathbf{case~} c \mathbf{~of~} 1
  "->"\widehat{s_1} \mid 2 "->" \widehat{s_2}$
\end{itemize}
For instance $\widehat{s_{`a}}(\All) = 2$ and $\widehat{s_{`a}}(\Bel) = 1$.

We define an equivalence $s=_g s'$ among finite strategy profiles, which we read as \emph{``$s$ and $s'$
\emph{have the same (underlying) game}''}.
\begin{definition}
  We say that two strategy profiles $s$ and $s'$ \emph{have the same game} and we
  write $s=_g s'$ iff by induction
  \begin{itemize}
  \item either $s=\prof{u}$ and $s'=\prof{u}$
  \item or $s=\prof{a,c,s_1,s_2}$ and $s'=\prof{a',c',s_1',s_2'}$ and $a=a'$,
    $s_1=_g s_1$ and $s_2=_g s_2'$.
  \end{itemize}
\end{definition}

We can define a family of finite strategy profiles that are of interest for decisions.
First we start with \emph{backward induction}.  Following~\cite{vestergaard06:IPL},
we consider `backward induction', not as a reasoning method, but as a predicate that
specifies some strategy profiles.
\begin{definition}[Backward induction]
  A finite strategy profile $s$ is \emph{backward induction} if it satisfies the predicate
  $\mathsf{BI}$, where $\mathsf{BI}$ is defined recursively as follows:
  \begin{itemize}
  \item $\mathsf{BI}(\prof{u})$, i.e., by definition an ending position is `backward induction'.
  \item $\mathsf{BI}(\prof{a, 1, s_1, s_2}) \ "<=>"\ \mathsf{BI}(s_1) \wedge \mathsf{BI}(s_2) \wedge \widehat{s_1}
    \ge \widehat{s_2}$.
  \item $\mathsf{BI}(\prof{a, 2, s_1, s_2}) \ "<=>"\ \mathsf{BI}(s_1) \wedge
    \mathsf{BI}(s_2) \wedge \widehat{s_2} \ge \widehat{s_1}$.
  \end{itemize}
\end{definition}

In other words, a strategy profile which is not an ending position is `backward
induction' if both its direct strategy subprofiles are and if the choice leads to a
better utility, as shown by the comparison of the utility assignments to the direct
strategy subprofiles.  The two following strategy profiles are `backward
induction'~\cite{osborne04a}(Example~158.1)
\begin{displaymath}
           \xymatrix@C=10pt{
          &\ar@{.>}[r]& *++[o][F]{\Al} \ar@/^1pc/[rr]^2 \ar@[blue]@{=>}@/^/[d]^1
          &&*++[o][F]{\Be} \ar@/^1pc/[rr]^2 \ar@[blue]@{=>}@/^/[d]^1  && 2,1\\
          &&1,2&&0,1
        }
\qquad
           \xymatrix@C=10pt{
          &\ar@{.>}[r]& *++[o][F]{\Al} \ar@[blue]@{=>}@/^1pc/[rr]^2 \ar@/^/[d]^1
          &&*++[o][F]{\Be} \ar@[blue]@{=>}@/^1pc/[rr]^2 \ar@{->}@/^/[d]^1  && 2,1\\
          &&1,2&&0,1
        }
\end{displaymath}

An agent is rational if she makes a choice dictated by backward induction and if she
keeps being rational in the future. We write this predicate $\Ratf$ where the index
$f$ insists on finiteness making it distinct from the predicate $\Ratoo$ on infinite
strategy profiles.

\begin{definition}[Rationality for finite strategy profiles]\label{def:rat}
The predicate $\Ratf$ is defined recursively as follows:
\begin{itemize}
\item  $\Ratf(\prof{u})$,
\item $\Ratf(\prof{a, c, s_1, s_2}) "<=>" `E \prof{a, c, s_1', s_2'}`:  \mathsf{StratProf},$~\\
 \hspace*{.5\textwidth}\parbox{.5\textwidth}{\begin{itemize}
  \item $\prof{a, c, s_1', s_2'} =_g\prof{a, c, s_1, s_2}$
  \item $\mathsf{BI}(\prof{a, c, s_1', s_2'})$
  \item $\Ratf(s_c)$
  \end{itemize}}
\end{itemize}
\end{definition}

Then we can  state a variant of Aumann theorem~\cite{aumann95} saying that backward
induction coincides with rationality.
\begin{theorem}
  \begin{math}
  `A s`:\mathsf{StratProf}, \Ratf(s) "<=>" \mathsf{BI}(s).
\end{math}
\end{theorem}

\section{Decisions in Infinite Strategy Profiles}
\label{sec:inf_Strat_Prof}

We extend the concept of backward induction and the concept of rationality to
infinite strategy profiles.  For that, we replace induction by
coinduction.\footnote{For readers not familiar with coinduction and not willing
  to read~\cite{jacobs12:_introd_coalg} or~\cite{DBLP:journals/tcs/Rutten00}, we advise her to
  pretend just that corecursive definitions define infinite objects and coinduction
  allows reasoning specifically on their infinite aspects, whereas recursive definition define
  finite objects and induction allows reasoning on their finite aspects.}  Notice that we mix up
recursive and corecursive definitions, and that we reason sometime by induction and
sometime by coinduction.  Therefore we advise the reader to be cautious and to pay
attention to when we use one or the other.
\ifLONG
\footnote{Notice that not all the authors
  are as cautious. For instance, Hargreaves-Heap and Varoufakis write
  (\cite{shaun04:_game_theor} p.27) ``\emph{The idea [of common knowledge] reminds
    one what happens when a camera is pointing to a television screen that conveys the
    same image recorded by the same camera:} an infinite self-reflection''.  Indeed
  \emph{common knowledge} is typically inductive whereas \emph{infinite
      self-reflection} is typically coinductive.}
\fi
We write $\mathsf{InfStratProf}$
the set of finite or infinite strategy profiles.
\begin{definition}
  The set \emph{finite or infinite strategy profiles} $\mathsf{InfStratProf}$ is
  defined \emph{corecursively} as follows:
  \begin{itemize}
  \item either given a \emph{utility assignment} $u$, then $\prof{u}`:
    \mathsf{InfStratProf}$, which corresponds to an ending position.
  \item or given an \emph{agent} $a$, a \emph{choice} $c$ and two \emph{strategy
      profiles} $s_1`:\mathsf{InfStratProf}$ and $s_2`:\mathsf{InfStratProf}$, then
    $\prof{a, c, s_1, s_2}`: \mathsf{InfStratProf}$.
  \end{itemize}
\end{definition}

We cannot define the utility assignments on all infinite strategy profiles, only on
those on which the utility can be ``computed''. The strategy profiles on which
utility assignments are defined are called \emph{convergent}, since when one follows
the path indicated by the choices one ``converges'', that is that one gets to an
ending position, i.e., a position where utilities are actually attributed. The
predicate \emph{convergent} is defined by induction, meaning that, on $s$, after
finitely many steps following the choices of $s$ an ending position is reached.
``Finitely many steps'' is a finite aspect and this is why we use an inductive
definition.
\begin{definition}[Convergent]
  Saying that $s$ is \emph{convergent} is written $\conv{s}$.  $\conv{s}$ is defined
  by induction as follows:
  \begin{itemize}
  \item $\conv{\prof{u}}$ or
  \item if $\conv{s_1}$ then  $\conv{\prof{a,1, s_1, s_2}}$ or
  \item if $\conv{s_2}$ then  $\conv{\prof{a,2, s_1, s_2}}$ or
  \end{itemize}
\end{definition}

On convergent strategy profiles we can assign utilities. The resulting function is
written $\hat{s}$ when applied to a strategy profile $s$.
\begin{definition}[Utility assignment]
$\hat{s}$ is defined corecursively on every strategy profile.
\begin{displaymath}
\begin{array}{lll}
\textit{~when~} &s = \prof{u}           &\quad \widehat{s} \quad =\quad f\\
\textit{~when~} & s = \prof{a,1,s_1,s_2} &\quad \widehat{s} \quad =  \quad \widehat{s_1} \qquad \\
\textit{~when~} & s = \prof{a,2,s_1,s_2} & \quad\widehat{s} \quad =  \quad\widehat{s_2} \qquad
\end{array}
\end{displaymath}
\end{definition}
The function $\hat{`.'}$ has to be specified on an infinite object and this is why we
use a corecursive definition.
\begin{proposition}
  If  $\conv{s}$, then $\hat{s}$ returns a value.
\end{proposition}
Actually convergent strategy profiles are not enough as we need to know the utility
assignment not only on the whole strategy profile but also on strategy subprofiles.
For that, we need to insure that from any internal position we can reach an ending
position, which yields that on any position we can assign a utility.  We call
\emph{always-convergent} such a predicate\footnote{Traditionally $\Box$ is the
  notation for the modality (i.e., the predicate transformer) \emph{always}.} and we
write it $\Conv{}$.
\begin{definition}[Always-convergent]~\label{def:al-conv}
  \begin{itemize}
  \item $\Conv{\prof{u}}$ that is that for whatever $u$, $\prof{u}$ is always-convergent
  \item $\Conv{\prof{a,c, s_1, s_2}}$ if
    \begin{itemize}
    \item $\prof{a,c, s_1, s_2}$ is convergent (i.e., $\conv{\prof{a,c, s_1, s_2}}$),
      and
    \item $s_1$ is always-convergent (i.e., $\Conv{s_1}$), and
    \item $s_2$ is always-convergent (i.e., $\Conv{s_2}$).
    \end{itemize}
  \end{itemize}
\end{definition}

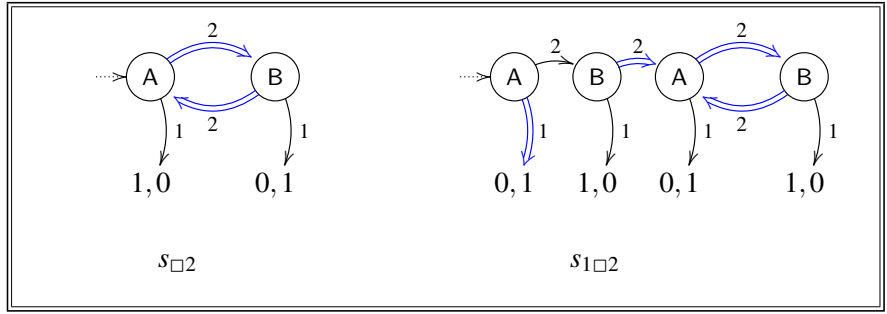
\begin{figure}[htb!]
  \centering
  \doublebox{\parbox{\iffullpage .65\textwidth\else .85\textwidth\fi}{ \begin{center} \begin{math}
     \begin{array}{ccc}
         \xymatrix@C=10pt{
          &\ar@{.>}[r]& *++[o][F]{\Al} \ar@[blue]@{=>}@/^1pc/[rr]^2 \ar@/^/[d]^1
          &&*++[o][F]{\Be} \ar@[blue]@{=>}@/^1pc/[ll]^2 \ar@/^/[d]^1 \\
          &&1,0&&0,1
        }
        &\qquad
        \xymatrix@C=10pt{
          &\ar@{.>}[r]& *++[o][F]{\Al} \ar@/^/[r]^2 \ar@[blue]@{=>}@/^/[d]^1
          &*++[o][F]{\Be} \ar@[blue]@{=>}@/^/[r]^2 \ar@/^/[d]^1
          & *++[o][F]{\Al} \ar@[blue]@{=>}@/^1pc/[rr]^2 \ar@/^/[d]^1
          &&*++[o][F]{\Be} \ar@[blue]@{=>}@/^1pc/[ll]^2 \ar@/^/[d]^1 \\
          &&0,1&1,0&0,1&&1,0
        }\\\\
        s_{\Box 2}  & s_{1 \Box 2}
      \end{array}
\end{math}
\end{center}
}
}
\caption{Two examples of strategy profiles}
\label{fig:alwr}
\end{figure}

\begin{proposition}
  $\Conv{s} \quad "=>" \quad \conv{s}.$
\end{proposition}
$s_{\Box 2}$ in Figure~\ref{fig:alwr} is a typically non convergent strategy profile,
wherever $s_{1 2 \Box 2}$ in the same figure is a typically convergent and not
always-convergent strategy profile.

Using the concept of always-convergence we can generalize the notion of backward
induction to this that the tradition calls \emph{subgame perfect
  equilibrium}~\cite{selten65:_spiel_behan_eines_oligop_mit} and which we write here
\textsf{SPE}.  In short \textsf{SPE} is a corecursive generalization of \textsf{BI}.
First we define an auxiliary predicate.
\begin{definition}[\textsf{PE}]
  \[
\begin{array}{ll}
\mathsf{PE}(s) \quad "<=>" \quad \Conv{s}\  &\wedge\ s = \prof{a, 1, s_1, s_2} "=>"
\widehat{s_1}(a) \ge \widehat{s_2}(a)\\
& \wedge\ s = \prof{a, 2, s_1, s_2} "=>"
\widehat{s_2}(a) \ge \widehat{s_1}(a)
\end{array}
\] %
\end{definition}
We define \emph{SPE} as \emph{always}-\textsf{PE}.  In other words, a strategy
profile $s$ is a subgame perfect equilibrium if $\Box\mathsf{PE}(s)$.
 $\Box$ applies to a predicate.
 \begin{definition}[Always]~\label{def:box}
   Given a predicate $P$, the predicate $\Box P$ is defined corecursively as follows.
   \begin{itemize}
   \item if $P(\prof{u})$ then $\Box P(\prof{u})$ and
   \item if $\Box P(s_1)$, $\Box P(s_2)$ and $P(\prof{a,c,s_1,s_2})$ then $\Box P(\prof{a,c,s_1,s_2})$
   \end{itemize}
 \end{definition}
 Formally \emph{SPE} is $\Box \mathsf{PE}$.  Besides we may notice that the notation
 used for always-convergence (Definition \ref{def:al-conv}) is consistent with
 Definition~\ref{def:box}.  Now thanks to \textsf {SPE} we can give a notion of
 rationality for infinite strategy profiles.  Like for finite strategy profiles we
 define \underline{corecursively}, this time, an equivalence $s=_g s'$ on infinite
 strategy profiles (read $s$ and $s'$ \emph{have the same game}).  Two strategy
 profiles have the same game if at each step, they have the same agent and their
 respective direct strategy subprofiles have the same game and only the choices differ.
\begin{definition}
  We say that two strategy profiles $s$ and $s'$ \emph{have the same game} and we
  write $s=_g s'$ iff corecursively
  \begin{itemize}
  \item either $s=\prof{u}$ and $s'=\prof{u}$
  \item or $s=\prof{a,c,s_1,s_2}$ and $s'=\prof{a',c',s_1',s_2'}$ and $a=a'$,
    $s_1=_g s_1$ and $s_2=_g s_2'$.
  \end{itemize}
\end{definition}
\begin{definition}[Rationality for finite or infinite strategy profiles]
  $\Ratoo$ is defined \emph{corecursively} as follows.
\begin{itemize}
\item  $\Ratoo(\prof{u})$,
\item $\Ratoo(\prof{a, c, s_1, s_2}) "<=>"
 `E \prof{a, c, s_1', s_2'}`:  \mathsf{InfStratProf}, \\
\hspace*{.2\textwidth}\prof{a, c, s_1', s_2'}  =_g\prof{a, c, s_1, s_2} \wedge \mathsf{SPE}(\prof{a, c, s_1', s_2'}) \wedge
  \Ratoo(s_c)$
\end{itemize}
\end{definition}
The reader may notice the similarity with Definition~\ref{def:rat} of rationality for
finite games. The difference is twofold: the definition is corecursive instead of
recursive and \textsf{BI} has been replaced by \textsf{SPE}.  Let us now define a
predicate that states the opposite of convergence~\footnote{People used to coinduction
  know why it is better to define \emph{divergence} directly instead of defining it
  as the negation of convergence.}
\begin{definition}[Divergence]
  $\diverg{s}$ is defined corecursively as follows:
  \begin{itemize}
  \item if $\diverg{s_1}$ then $\diverg{\prof{a,1,s_1,s_2}}$,
  \item if $\diverg{s_2}$ then $\diverg{\prof{a,2,s_1,s_2}}$.
  \end{itemize}
\end{definition}
$s_{\Box 2}$ in Figure~\ref{fig:alwr} is a typical divergent strategy profile.  The
main theorem of this paper can then be stated, saying that there exists a strategy
profile that is both divergent and rational.

\medskip

\doublebox{\parbox{.85\textwidth}{
  \begin{theorem}[Risk of divergence]
    $`E s`:\mathsf{InfStratProf},  \Ratoo(s) \  \wedge\ \diverg{s}$.
  \end{theorem}
}}

\ifLONG
\medskip

Witnesses of divergent and rational strategy profiles will be given in
Section~\ref{sec:ompede} and
Section~\ref{sec:twoex}.
\fi

\section{Extrapolating the centipede}
\label{sec:oopede}

As an illustration of the above concepts, we show, in this section, two simple
extensions to infinity of a folklore example.  The centipede has been proposed by
Rosenthal~\cite{rosenthal81:_games_of_perfec_infor_predat}.  Starting from a wording
suggested by Aumann~\cite{aumann95} we study two infinite generalization\footnote{The
  reason why we call them \emph{\infpede} and
  \emph{\ompede}.}. Wikipedia~\cite{wiki:centipede} says:
\begin{quotation}
  Consider two players: Alice and Bob. Alice moves first. At the start of the game,
  Alice has two piles of coins in front of her: one pile contains 4 coins and the
  other pile contains 1 coin. Each player has two moves available: either \textbf{"take"} the
  larger pile of coins and give the smaller pile to the other player or \textbf{"push"} both
  piles across the table to the other player. Each time the piles of coins pass
  across the table, the quantity of coins in each pile doubles.
\end{quotation}

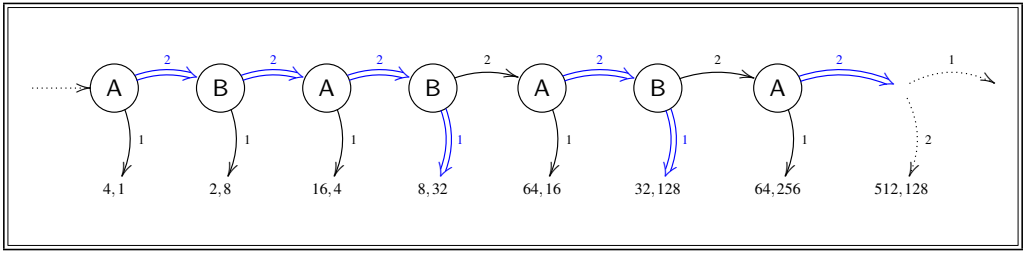
\begin{figure}[tbh!]
  \centering
  \doublebox{\parbox{\textwidth}{
  \iffullpage
   \begin{displaymath}
    \xymatrix@C=12pt{
      &\ar@{.>}[r]& \nodA\flr{r}&\nodB\flr{r}&\nodA\flr{r}&\nodB\fld{r}&\nodA\flr{r}&\nodB\fld{r}%
      &\nodA\flr{r} &\nodB\flr{r}&\nodA\flr{r} &\ar@{.>}@/^/[r]^r&\\
      &4,1&2,8&16,4&8,32&64,16&32,128&64,256&512,128&256,1024&&}
  \end{displaymath}
  \else
  \begin{tiny}
       \begin{displaymath}
    \xymatrix@C=22pt{
      \ar@{.>}[r]& \nodA\flr{r}&\nodB\flr{r}&\nodA\flr{r}&\nodB\fld{r}&\nodA\flr{r}&\nodB\fld{r}%
      &\nodA\flr{r} &\ar@{.>}@/^/[r]^1\ar@{.>}@/^/[d]^2&\\
       &4,1&2,8&16,4&8,32&64,16&32,128&64,256&512,128&&}
  \end{displaymath}
  \end{tiny}
  \fi
}}
  \caption{A sketch of a strategy profile of the \infpede{}.}
  \label{fig:ABAB}
\end{figure}

\subsection{The \infpede}
\label{sec:infpede}

$P_{\infty}(s)$ is a set of strategy profiles extending the strategy profiles of the
centipede.  Such an infinite strategy profile can only be is sketched on
Figure~\ref{fig:ABAB} .  Actually proposing an infinite extension of the centipede is
quite natural for two reasons.  First there is no natural way to make the game
finite.  Indeed in the definition of the game, nothing precise is said about its end,
when no player decides to take a pile. For instance, Wikipedia~\cite{wiki:centipede}
says:
\begin{quotation}
  The game continues for a fixed number of rounds or until a player decides to end
  the game by pocketing a pile of coins.
\end{quotation}
We do no know what the utilities are in the end position described as ``a fixed
number of rounds''.  Since $\All$ started, we can assume that the end after a fixed
number of rounds is $\Bel$'s turn and that there are outcomes like:
\begin{enumerate}
\item $\Bel$ receives $2^{n+1}$ coins and $\All$ receives $2^{n+3}$ coins like for
  the previous $\Bel$ rounds and that is all.
\item $\Bel$ chooses between
  \begin{enumerate}
  \item receiving $2^{n+1}$ coins whereas $\All$ receives $2^{n+3}$ or
  \item sharing with $\All$, each one receiving $2^{n+2}$.
  \end{enumerate}
\item Both $\All$ and $\Bel$ receive nothing.
\end{enumerate}
Moreover the statement \emph{``Each player has two moves available: either
  \textbf{``take''} ... or \textbf{push}...''} is not true, in the ending position.
We are not hair-splitting since the end positions are the initializations of the
(backward) induction and must be defined as precisely as the induction step. Ending
with 2.(b) does not produce the same backward induction as the others.
Let us consider the strategy profiles
\begin{eqnarray*}
p_n &=& \prof{\All, 1, \prof{\All"|->"2^{2n+2},\Bel"|->"2^{2n}},`p_n}\\
`p_n &=& \prof{\Bel, 1, \prof{\All"|->"2^{2n+1},\Bel"|->"2^{2n+3}}, p_{n+1}}
\end{eqnarray*}
In words, the $p_n$'s and the $`p_n$'s are the strategy subprofiles of the \infpede{}
in which Alice and Bob stop always.  Notice that
\begin{center}
  \begin{math}
    \widehat{p_n}(\All) = 2^{2n+2} \qquad  \widehat{p_n}(\Bel) = 2^{2n}
  \end{math}

  \begin{math}
    \widehat{`p_n}(\All) = 2^{2n+1} \qquad  \widehat{`p_n}(\Bel) = 2^{2n+3}
  \end{math}
\end{center}
\begin{theorem}~
  \begin{enumerate}
  \item $`A n`:\nat, \mathsf{SPE} (p_n) \wedge \mathsf{SPE}(`p_n)$,
  \item $`A s`:\mathsf{InfStratProf}, s =_g p_0 \wedge \mathsf{SPE}(s) \quad "<=>" s = p_0$.
  \end{enumerate}
 \end{theorem}
 In other words, all the $p_n$'s and the $`p_n$'s are `backward induction'. Moreover
 for the \infpede{}, $p_0$ is the only `backward induction'. strategy profile.
 \begin{proof}
One can easily prove that for all $n$, $\Conv{p_n}$ and  $\Conv{`p_n}$.

   Assuming $\mathsf{SPE} (`p_n)$ and $\mathsf{SPE} (p_{n+1})$ (coinduction) and since
   \begin{displaymath}
   \widehat{\prof{\All"|->"2^{2n+2},\Bel"|->"2^{2n}}(\All)\ge \widehat{`p_n}(\All)}
 \end{displaymath}
we conclude that
   \begin{math}
     \mathsf{SPE}(\prof{\All, 1, \prof{\All"|->"2^{2n+2},\Bel"|->"2^{2n}},`p_n})
   \end{math}
that is $\mathsf{SPE}(p_n)$.   The proof of $\mathsf{SPE}(`p_n)$ is similar.

For the proof of 2. we notice that in a  strategy profile in $\mathsf{SPE}$ with the same game
as $p_0$,
there is no strategy subprofile such that the agent chooses $2$ and the next agent
chooses $1$.  Assume the strategy subprofile is $s_n = \prof{\All,2, \prof{\All"|->"2^{2n+2},\Bel"|->"2^{2n}},
  \prof{\Bel,1,\prof{\All"|->"2^{2n+1},\Bel"|->"2^{2n+3}},`s_n}}$. ans that $\mathsf{SPE}(s_n)$ and $\mathsf{SPE}(`s_n)$.    If it would be
the case and if we write $t = \prof{\All"|->"2^{2n+2},\Bel"|->"2^{2n}}$ and
$t'=\prof{\Bel,1,\prof{\All"|->"2^{2n+1},\Bel"|->"2^{2n+3}},`s_n}$, we notice that
$\hat{t}(A)=2^{2n+2}> \widehat{t'}(A) = 2^{2n+1}$.  This is in contradiction with $\mathsf{SPE}(s_n)$.
 \end{proof}
We deduce that the
 strategy profile $d_0$, which diverges, is not in $\Ratoo$ and more generally there
 is no strategy profile in $\Ratoo$ for the \infpede.
\begin{eqnarray*}
d_n &=& \prof{\All, 2,\prof{\All"|->"2^{2n+2},\Bel"|->"2^{2n}},`p_n}\\
`d_n &=&  \prof{\Bel, 2, \prof{\All"|->"2^{2n+1},\Bel"|->"2^{2n+3}}, p_{n+1}}
\end{eqnarray*}
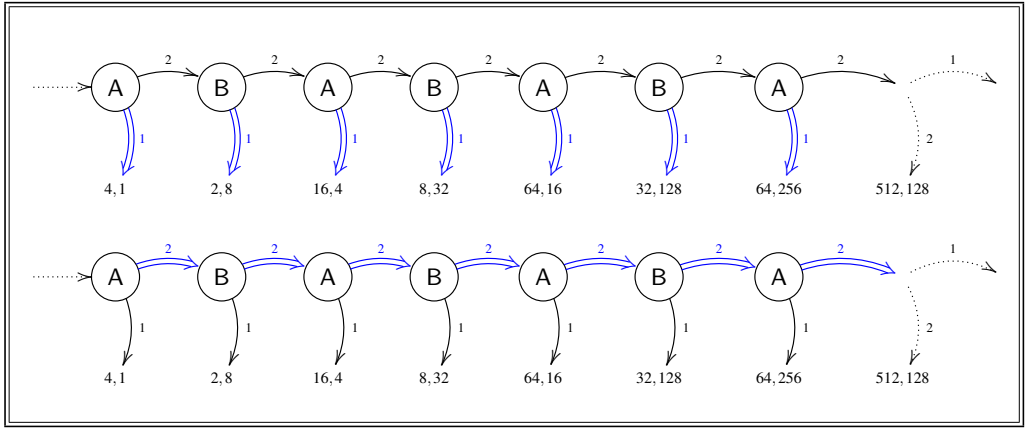
\begin{figure}[th!]
  \centering
  \doublebox{\parbox{\textwidth}{
  \iffullpage
   \begin{displaymath}
    \xymatrix{
      \nodA\fld{r}&\nodB\fld{r}&\nodA\fld{r}&\nodB\fld{r}&\nodA\fld{r}&\nodB\fld{r}%
      &\nodA\fld{r} &\nodB\fld{r}&\nodA\fld{r} &\ar@{.>}@/^/[r]^r&\\
      4,1&2,8&16,4&8,32&64,16&32,128&64,256&512,128&256,1024&&}
  \end{displaymath}
   \begin{displaymath}
    \xymatrix{
      \nodA\flr{r}&\nodB\flr{r}&\nodA\flr{r}&\nodB\flr{r}&\nodA\flr{r}&\nodB\flr{r}%
      &\nodA\flr{r} &\nodB\flr{r}&\nodA\flr{r} &\ar@{.>}@/^/[r]^r&\\
      4,1&2,8&16,4&8,32&64,16&32,128&64,256&512,128&256,1024&&}
  \end{displaymath}
  \else
  \begin{tiny}
       \begin{displaymath}
    \xymatrix@C=22pt{
      \ar@{.>}[r]& \nodA\fld{r}&\nodB\fld{r}&\nodA\fld{r}&\nodB\fld{r}&\nodA\fld{r}&\nodB\fld{r}%
      &\nodA\fld{r} &\ar@{.>}@/^/[r]^1\ar@{.>}@/^/[d]^2&\\
       &4,1&2,8&16,4&8,32&64,16&32,128&64,256&512,128&&}
  \end{displaymath}
       \begin{displaymath}
    \xymatrix@C=22pt{
     \ar@{.>}[r]&  \nodA\flr{r}&\nodB\flr{r}&\nodA\flr{r}&\nodB\flr{r}&\nodA\flr{r}&\nodB\flr{r}%
      &\nodA\flr{r} &\ar@{.>}@/^/[r]^1\ar@{.>}@/^/[d]^2&\\
      & 4,1&2,8&16,4&8,32&64,16&32,128&64,256&512,128&&}
  \end{displaymath}
  \end{tiny}
  \fi
}}
  \caption{Strategy profiles $p_0$ and $d_0$ of the \infpede{}.}
  \label{fig:infped}
\end{figure}

\subsection{The \ompede}
\label{sec:ompede}

We know\footnote{Usually agents do not believe this.  See~\cite{lescanne14:_intel} for
  a discussion of the beliefs of the agents w.r.t. the infiniteness of the world.} that \emph{``trees
  don't grow to the sky''}.  In our case this means that there is a natural number
$`w$ after which piles cannot be doubled.footnote{People speak of \emph{limited
    payroll}.}  In other words, after $`w$, the piles keep the same size $2^{`w}$.
An example of strategy profile is sketched on Figure~\ref{fig:omped}.  In this family
of strategy profiles, which we write $P_{`w}$ , the utilities stay stable after the
$`w^{th}$ positions.  Every \emph{always-convergent} strategy profile of $P_{`w}$,
such that agents \textbf{push} until $`w$ is in \textsf{SPE}.  We conclude the
existence of rational divergent strategy profiles in $P_{`w}$.  In other words in the
$\ompede$ there is a risk of divergence.
\begin{figure}[ht]
\newcommand{\oped}{%
\begin{displaymath}
 \xymatrix{
      \ar@{.>}[r]& \nodA\flr{r}&\nodB\flr{r}&\nodA\flr{r}&\nodB\flr{r}&\ar@[blue]@{:>}@/^/[r]%
      &\nodA\flr{r} &\nodB\fld{r}&\nodA\fld{r} &\\
      &4,1&2,8&16,4&8,32&&2^{`w},2^{`w}&2^{`w},2^{`w}&2^{`w},2^{`w}&&}%
\end{displaymath}
}
  \centering
  \doublebox{\parbox{\textwidth}{
  \iffullpage
   \oped
    \else
  \begin{tiny}
    \oped
  \end{tiny}
  \fi
}}
  \caption{A `backward induction` strategy profile for the \ompede{}.}
  \label{fig:omped}
\end{figure}
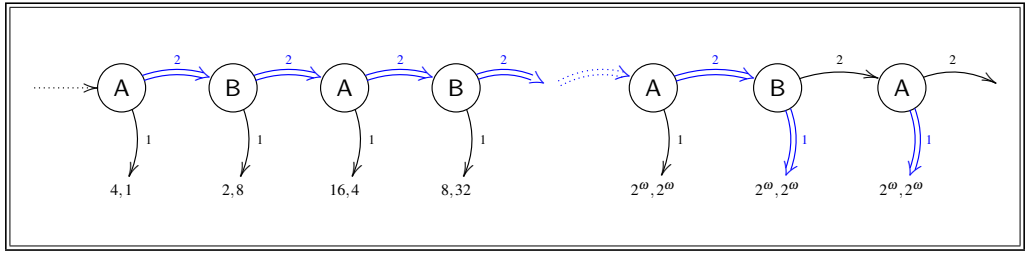
\begin{theorem}
  $`E s `: P_{`w}, \Ratoo(s) \wedge \diverg{s}.$
\end{theorem}
One may imagine that divergence is when optimistic agents hope a reverse of the
tendency.

\paragraph*{Comments:}
The \ompede{} example is degenerated, but it is interesting in two respects.  First,
it shows a very simple and naive case of rational divergence.  Second it shows that
cutting the infinite game, case 2. (b) is the most natural way, with a equilibrium in
which agents \textbf{take} until the end.

\section{Two examples}
\label{sec:twoex}

\paragraph{$0,1$ strategy profiles}
$0,1$ strategy profiles are strategy profiles with the shape of an infinite ``comb''
in which the utilities are $0$ for the agent who quits and $1$ for the other agent.
It can be shown~\cite{DBLP:conf/calco/Lescanne13} that strategy profiles where one
agent continues always and the other quits infinitely often (in other words the other
agent never continues always) are in \textsf{SPE}.  For this reason, the strategy profile where
both agents continue always is in $\Ratoo$, which shows that divergence is rational.

\paragraph{The dollar auction}
The \emph{dollar auction} is a well known
game~\cite{Shubik:1971,leininger89:_escal_and_cooop_in_confl_situat,oneill86:_inten_escal_and_dollar_auction}. Its
strategy profiles have the same infinite comb shape as the $0,1$ strategy
profiles, the \infpede{} and the \ompede{}
with the sequence of pairs of utilities:

\begin{footnotesize}
  \begin{center}
    \begin{math}
      (0,100)\ (95,0)\ (-5,95)\ (90,-5) \ (-10,90)\ (85,-10)\ \ldots\ (-5n,100-5n) \
      (100-5(n+1),-5n)\ \ldots
    \end{math}
  \end{center}
\end{footnotesize}
and corresponds to an auction in which the bet of the looser is not returned to her.
We have shown~\cite{DBLP:journals/acta/LescanneP12} that the dollar auction may
diverge with rational agents. People speak of \emph{escalation} in this case.  The
divergent strategy profile of the dollar auction is in $\Ratoo$.

\section{Reflection}
\label{sec:refl}
\ifLONG
\rightline{\parbox{4.8cm}{\begin{it} Je est un autre\end{it}
                        \begin{it} (I is another\end{it})}}%
\rightline{\emph{Arthur Rimbaud} (1854-1891) }

\medskip
\fi

Examples like the dollar auction or the $0,1$ raise the following question: ``How is
it possible in an escalation that the agents do not see that they are entering a
hopeless process?''.  The answer is ``reflection''.  Indeed, when reasoning, betting
and choosing, the agents should leave the world where they live and act in order to
observe the divergence.  If they are wise, they change their beliefs in an infinite
world as soon as they realize that they go nowhere~\cite{stanovich09:_two}.  This
ability is called reflection and is connected to observability, from the theoretical
computer science point of view, which is itself connected to coalgebras and to
coinduction~\cite{jacobs12:_introd_coalg}. In other words, agents should leave the
environment in which they are enclosed and observe themselves.\ifLONG Like the poet, they
should be able to claim \emph{``I is another''} whom I consider as an object.\fi

\section{Singularities and divergence}
\label{sec:sing}

Divergence is called singularity, bubble, crash, escalation, or turbulence according
to the context or the scientific field.  In mechanics this is considered as a
topics by itself.   Leonardo da Vinci's drawings \ifLONG (Fig.~\ref{fig:turbh} left)\fi show that he considered early turbulence and
vortices and only Reynolds during the XIX$^{\textrm{th}}$ century studied it from a
scientific point of view.  In many other domains, phenomena of this family are
rejected from the core of the field, despite they have been observed experimentally.
Scientists, among them mainstream economists~\cite{colander04:_chang_face_mains_econom}, prefer
smoothness, continuity and equilibria~\cite{Neoclass_Ecoc} and they often claim that
departing from this leads to ``paradoxes''~\cite{Shubik:1971}.
In~\cite{DBLP:journals/corr/abs-1112-1185}, we surveyed Zeno of Elea's paradox from
the point of view of coinduction, as well as Weierstrass
function~\cite{weierstrass72}, the first mathematical example showing discontinuity
at the infinite.  Here we would like to address two other cases.
\ifLONG
\begin{figure}[htb!]
  \centering
  \includegraphics[width=0.25\columnwidth]{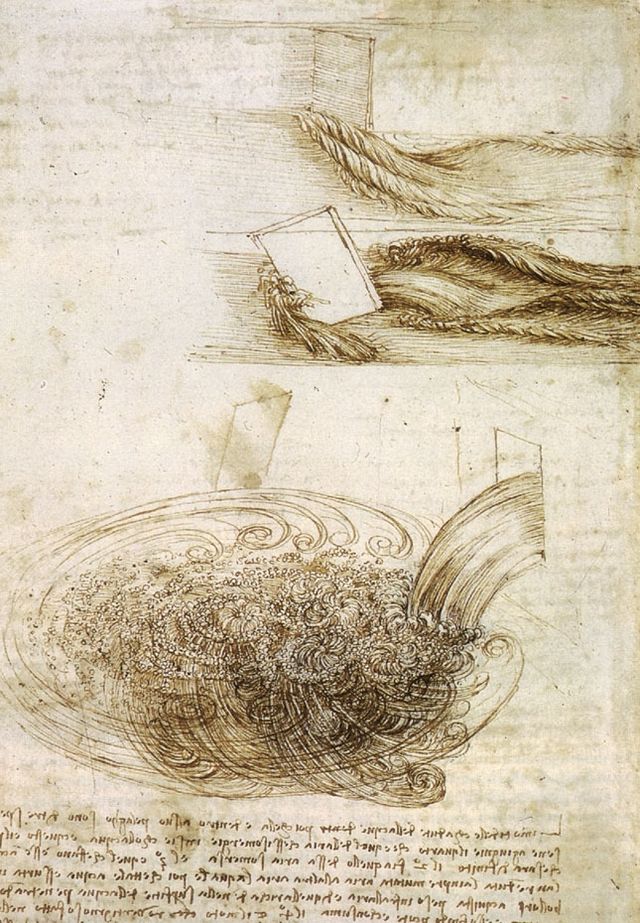}
\qquad\qquad
\includegraphics[height=0.35\columnwidth]{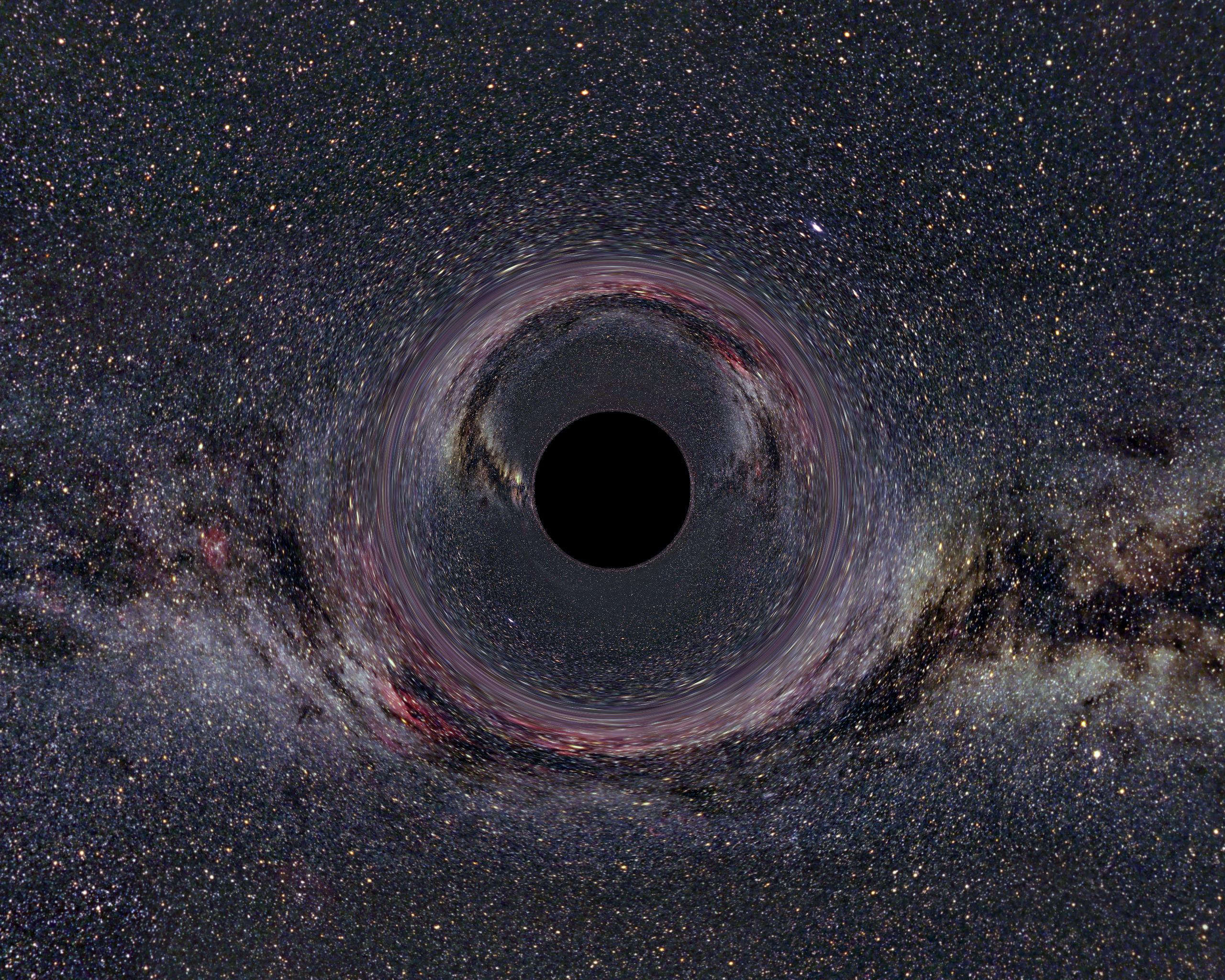}
  \caption{Da Vinci's drawings  (left) and a artist view of a blackhole  (right)
    {\tiny Wikimedia commons}}
  \label{fig:turbh}
\end{figure}
\fi
In 1935, that is one year before his famous article in the \emph{Proceedings of the
  London Mathematical Society}~\cite{Turing:1936:CNA}, Alan Turing wrote a
paper~\cite{turing35:_first_comput_number_compt_rendues} presenting his result for a
publication in the Proceedings of the French Academy of Science.  In this paper he
calls ``nasty'' a machine that terminates and ``nice'' a machine that does not
terminate, showing his positive view of non terminating computations.\footnote{Notice
  that he changed his terminology
  in~\cite{turing35:_first_comput_number_compt_rendues} and calls ``circular'' the
  terminating machine and ``circular-free'' the non terminating machine.}
In 1795, Laplace published his book \emph{Exposition du Syst\`{e}me du Monde} and
proposed the first clear vision of the notion of blackhole\ifLONG (Fig.~\ref{fig:turbh} right)\fi, but probably in order not
to hurt his contemporaries, he found wiser to remove this presentation from the third
edition of his book.  Then we had to wait Schwartzschild in 1915, few months after
the publication by Einstein of the general theory of relativity, for a second
proposal of the concept of blackhole.  But at that time the general relativity was
not yet fully accepted as were not blackholes.  Only recently, at the end of the last
century, the general relativity has been considered as ``the'' theory of gravitation and
there is no more doubt on the existence of blackholes. Since blackholes are
singularities in gravitation, they are for the general theory of relativity the
equivalent of divergent strategy profiles for game theory.

\subsection*{Contribution of this paper}

Unlike previous presentations of similar
results~\cite{DBLP:journals/acta/LescanneP12,DBLP:conf/calco/Lescanne13,} here we
focus on the concept of strategy profile which is central for the those of
convergence, of divergence and of equilibrium and is more targeted for a workshop on
\emph{strategy reasoning}.  Moreover we introduce the \emph{\ompede{}} (a new
infinite version of the centipede) and \emph{``divergent'' strategy profiles} are
those that where called ``escalation'' in previous literature. This terminology seems
better fitted for its duality with convergence.

\section{Conclusion}
\label{sec:conclusion}

We have shown that strategy profiles in which no fixed limit is set must be studied
as infinite objects using coinduction and corecursion.  In these infinite objects,
the risk of divergence is real and should be  considered seriously.

\nocite{DBLP:conf/mfcs/HonsellLR12}

\begin{thebibliography}{10}
\providecommand{\bibitemdeclare}[2]{}
\providecommand{\surnamestart}{}
\providecommand{\surnameend}{}
\providecommand{\urlprefix}{Available at }
\providecommand{\url}[1]{\texttt{#1}}
\providecommand{\href}[2]{\texttt{#2}}
\providecommand{\urlalt}[2]{\href{#1}{#2}}
\providecommand{\doi}[1]{doi:\urlalt{http://dx.doi.org/#1}{#1}}
\providecommand{\bibinfo}[2]{#2}

\bibitemdeclare{article}{aumann95}
\bibitem{aumann95}
\bibinfo{author}{Robert~J. \surnamestart Aumann\surnameend}
  (\bibinfo{year}{1995}): \emph{\bibinfo{title}{Backward Induction and Common
  Knowledge of Rationality}}.
\newblock {\sl \bibinfo{journal}{Games and Economic Behavior}}
  \bibinfo{volume}{8}, pp. \bibinfo{pages}{6--19}.

\bibitemdeclare{article}{colander04:_chang_face_mains_econom}
\bibitem{colander04:_chang_face_mains_econom}
\bibinfo{author}{David \surnamestart Colander\surnameend},
  \bibinfo{author}{Richard P.~F. \surnamestart Holt\surnameend} \&
  \bibinfo{author}{Jr~\surnamestart Rosser\surnameend, Barkley~J.}
  (\bibinfo{year}{2004}): \emph{\bibinfo{title}{The Changing Face of Mainstream
  Economics}}.
\newblock {\sl \bibinfo{journal}{Review of Political Economy}}
  \bibinfo{volume}{16}(\bibinfo{number}{4}), p. \bibinfo{pages}{485–499},
  \doi{10.1080/0953825042000256702}.

\bibitemdeclare{book}{shaun04:_game_theor}
\bibitem{shaun04:_game_theor}
\bibinfo{author}{Shaun \surnamestart Hargreaves-Heap\surnameend} \&
  \bibinfo{author}{Yanis \surnamestart Varoufakis\surnameend}
  (\bibinfo{year}{2004}): \emph{\bibinfo{title}{Game Theory: A Critical
  Introduction}}.
\newblock \bibinfo{publisher}{Routledge}.

\bibitemdeclare{inproceedings}{DBLP:conf/mfcs/HonsellLR12}
\bibitem{DBLP:conf/mfcs/HonsellLR12}
\bibinfo{author}{Furio \surnamestart Honsell\surnameend},
  \bibinfo{author}{Marina \surnamestart Lenisa\surnameend} \&
  \bibinfo{author}{Rekha \surnamestart Redamalla\surnameend}
  (\bibinfo{year}{2012}): \emph{\bibinfo{title}{Categories of Coalgebraic
  Games}}.
\newblock In \bibinfo{editor}{Branislav \surnamestart Rovan\surnameend},
  \bibinfo{editor}{Vladimiro \surnamestart Sassone\surnameend} \&
  \bibinfo{editor}{Peter \surnamestart Widmayer\surnameend}, editors: {\sl
  \bibinfo{booktitle}{Mathematical Foundations of Computer Science 2012 - 37th
  International Symposium, {MFCS} 2012, Bratislava, Slovakia, August 27-31,
  2012. Proceedings}}, {\sl \bibinfo{series}{Lecture Notes in Computer
  Science}} \bibinfo{volume}{7464}, \bibinfo{publisher}{Springer}, pp.
  \bibinfo{pages}{503--515}, \doi{10.1007/978-3-642-32589-2\_45}.
\newblock \urlprefix\url{http://dx.doi.org/10.1007/978-3-642-32589-2_45}.

\bibitemdeclare{book}{jacobs12:_introd_coalg}
\bibitem{jacobs12:_introd_coalg}
\bibinfo{author}{Bart \surnamestart Jacobs\surnameend} (\bibinfo{year}{2012}):
  \emph{\bibinfo{title}{Introduction to Coalgebra. Towards Mathematics of
  States and Observations}}.
\newblock \bibinfo{publisher}{Online book}.
\newblock \bibinfo{note}{Version 2.0}.

\bibitemdeclare{article}{leininger89:_escal_and_cooop_in_confl_situat}
\bibitem{leininger89:_escal_and_cooop_in_confl_situat}
\bibinfo{author}{Wolfgang \surnamestart Leininger\surnameend}
  (\bibinfo{year}{1989}): \emph{\bibinfo{title}{Escalation and cooperation in
  conflict situations}}.
\newblock {\sl \bibinfo{journal}{J. of Conflict Resolution}}
  \bibinfo{volume}{33}, pp. \bibinfo{pages}{231--254}.

\bibitemdeclare{article}{DBLP:journals/corr/abs-1112-1185}
\bibitem{DBLP:journals/corr/abs-1112-1185}
\bibinfo{author}{Pierre \surnamestart Lescanne\surnameend}
  (\bibinfo{year}{2011}): \emph{\bibinfo{title}{Rationality and Escalation in
  Infinite Extensive Games}}.
\newblock {\sl \bibinfo{journal}{CoRR}} \bibinfo{volume}{abs/1112.1185}.
\newblock \urlprefix\url{http://arxiv.org/abs/1112.1185}.

\bibitemdeclare{inproceedings}{DBLP:conf/calco/Lescanne13}
\bibitem{DBLP:conf/calco/Lescanne13}
\bibinfo{author}{Pierre \surnamestart Lescanne\surnameend}
  (\bibinfo{year}{2013}): \emph{\bibinfo{title}{A Simple Case of Rationality of
  Escalation}}.
\newblock In \bibinfo{editor}{Reiko \surnamestart Heckel\surnameend} \&
  \bibinfo{editor}{Stefan \surnamestart Milius\surnameend}, editors: {\sl
  \bibinfo{booktitle}{CALCO}}, {\sl \bibinfo{series}{Lecture Notes in Computer
  Science}} \bibinfo{volume}{8089}, \bibinfo{publisher}{Springer}, pp.
  \bibinfo{pages}{191--204}.
\newblock \urlprefix\url{http://dx.doi.org/10.1007/978-3-642-40206-7_15}.

\bibitemdeclare{techreport}{lescanne14:_intel}
\bibitem{lescanne14:_intel}
\bibinfo{author}{Pierre \surnamestart Lescanne\surnameend}
  (\bibinfo{year}{2014}): \emph{\bibinfo{title}{Intelligent escalation and the
  principle of relativity}}.
\newblock \bibinfo{type}{Technical Report} \bibinfo{number}{ensl-01096264v4},
  \bibinfo{institution}{LIP - ENS Lyon.}

\bibitemdeclare{article}{DBLP:journals/acta/LescanneP12}
\bibitem{DBLP:journals/acta/LescanneP12}
\bibinfo{author}{Pierre \surnamestart Lescanne\surnameend} \&
  \bibinfo{author}{Matthieu \surnamestart Perrinel\surnameend}
  (\bibinfo{year}{2012}): \emph{\bibinfo{title}{"{Backward}" coinduction,
  {Nash} equilibrium and the rationality of escalation}}.
\newblock {\sl \bibinfo{journal}{Acta Inf.}}
  \bibinfo{volume}{49}(\bibinfo{number}{3}), pp. \bibinfo{pages}{117--137}.
\newblock \urlprefix\url{http://dx.doi.org/10.1007/s00236-012-0153-3}.

\bibitemdeclare{article}{oneill86:_inten_escal_and_dollar_auction}
\bibitem{oneill86:_inten_escal_and_dollar_auction}
\bibinfo{author}{Barry \surnamestart O'Neill\surnameend}
  (\bibinfo{year}{1986}): \emph{\bibinfo{title}{International escalation and
  the dollar auction}}.
\newblock {\sl \bibinfo{journal}{J. of Conflict Resolution}}
  \bibinfo{volume}{30}(\bibinfo{number}{33-50}).

\bibitemdeclare{book}{osborne04a}
\bibitem{osborne04a}
\bibinfo{author}{Martin~J. \surnamestart Osborne\surnameend}
  (\bibinfo{year}{2004}): \emph{\bibinfo{title}{An Introduction to Game
  Theory}}.
\newblock \bibinfo{publisher}{Oxford}.

\bibitemdeclare{article}{rosenthal81:_games_of_perfec_infor_predat}
\bibitem{rosenthal81:_games_of_perfec_infor_predat}
\bibinfo{author}{R.~W. \surnamestart Rosenthal\surnameend}
  (\bibinfo{year}{1981}): \emph{\bibinfo{title}{Games of perfect information,
  predatory pricing and the chain-store paradox}}.
\newblock {\sl \bibinfo{journal}{Journal of Economy Theory}}
  \bibinfo{volume}{25}(\bibinfo{number}{1}), pp. \bibinfo{pages}{92--100}.

\bibitemdeclare{article}{DBLP:journals/tcs/Rutten00}
\bibitem{DBLP:journals/tcs/Rutten00}
\bibinfo{author}{Jan J. M.~M. \surnamestart Rutten\surnameend}
  (\bibinfo{year}{2000}): \emph{\bibinfo{title}{Universal coalgebra: a theory
  of systems}}.
\newblock {\sl \bibinfo{journal}{Theor. Comput. Sci.}}
  \bibinfo{volume}{249}(\bibinfo{number}{1}), pp. \bibinfo{pages}{3--80}.
\newblock \urlprefix\url{http://dx.doi.org/10.1016/S0304-3975(00)00056-6}.

\bibitemdeclare{article}{selten65:_spiel_behan_eines_oligop_mit}
\bibitem{selten65:_spiel_behan_eines_oligop_mit}
\bibinfo{author}{Reinhard \surnamestart Selten\surnameend}
  (\bibinfo{year}{1965}): \emph{\bibinfo{title}{{Spieltheoretische Behandlung
  eines Oligopolmodells mit Nachfragetr\"agheit}}}.
\newblock {\sl \bibinfo{journal}{Zeitschrift f\"ur gesamte Staatswissenschaft}}
  \bibinfo{volume}{121}.

\bibitemdeclare{article}{Shubik:1971}
\bibitem{Shubik:1971}
\bibinfo{author}{Martin \surnamestart Shubik\surnameend}
  (\bibinfo{year}{1971}): \emph{\bibinfo{title}{The Dollar auction game: A
  paradox in noncooperative behavior and escalation}}.
\newblock {\sl \bibinfo{journal}{Journal of Conflict Resolution}}
  \bibinfo{volume}{15}(\bibinfo{number}{1}), pp. \bibinfo{pages}{109--111}.

\bibitemdeclare{inbook}{stanovich09:_two}
\bibitem{stanovich09:_two}
\bibinfo{author}{Keith~E. \surnamestart Stanovich\surnameend}
  (\bibinfo{year}{2009}): \emph{\bibinfo{title}{Two minds: {Dual} processis and
  beyond, J. Evans and K. Frankish eds.}}, chapter
  \bibinfo{chapter}{Distinguishing the reflective, algorithmic, and autonomous
  minds: Is it time for a tri-process theory?}
\newblock \bibinfo{publisher}{Oxford University Press}.

\bibitemdeclare{misc}{turing35:_first_comput_number_compt_rendues}
\bibitem{turing35:_first_comput_number_compt_rendues}
\bibinfo{author}{A.~M. \surnamestart Turing\surnameend} (\bibinfo{year}{1935}):
  \emph{\bibinfo{title}{First draft of pr\'{e}cis of 'Computable Numbers', made
  for Comptes Rendues.}}
\newblock
  \bibinfo{howpublished}{\url{http://www.turingarchive.org/browse.php/K/4}}.

\bibitemdeclare{article}{Turing:1936:CNA}
\bibitem{Turing:1936:CNA}
\bibinfo{author}{A.~M. \surnamestart Turing\surnameend} (\bibinfo{year}{1936}):
  \emph{\bibinfo{title}{On Computable Numbers, with an Application to the
  {Entscheidungsproblem}}}.
\newblock {\sl \bibinfo{journal}{Proceedings of the London Mathematical
  Society. Second Series}} \bibinfo{volume}{42}, pp. \bibinfo{pages}{230--265}.

\bibitemdeclare{article}{vestergaard06:IPL}
\bibitem{vestergaard06:IPL}
\bibinfo{author}{Ren{\'e} \surnamestart Vestergaard\surnameend}
  (\bibinfo{year}{2006}): \emph{\bibinfo{title}{A Constructive Approach to
  Sequential {Nash} Equilibria}}.
\newblock {\sl \bibinfo{journal}{Inf. Process. Lett.}} \bibinfo{volume}{97},
  pp. \bibinfo{pages}{46--51}.

\bibitemdeclare{article}{}
\bibitem{}
\bibinfo{author}{Ren{\'e} \surnamestart Vestergaard\surnameend}
  (\bibinfo{year}{2007}): \emph{\bibinfo{title}{Cascaded Games}}
  \bibinfo{volume}{4545}, pp. \bibinfo{pages}{185--201}.
\newblock \urlprefix\url{http://dx.doi.org/10.1007/978-3-540-73433-8_14}.

\bibitemdeclare{article}{weierstrass72}
\bibitem{weierstrass72}
\bibinfo{author}{Karl \surnamestart Weierstrass\surnameend}
  (\bibinfo{year}{1872}): \emph{\bibinfo{title}{{\"Uber} continuirliche
  {Funktionen} eines reellen {Arguments}, die f\"ur keinen {Werth} des
  letzteren einen bestimmten {Differentialquotienten} besitzen}}.
\newblock {\sl \bibinfo{journal}{\textrm{in} \textit{Karl Weiertrass
  Mathematische Werke, {Abhandlungen II}} Gelesen in der K\"onigl. {Akademie}
  der {Wisseschaften}, am 18 Juli 1872}}, pp. \bibinfo{pages}{71--74}.

\bibitemdeclare{book}{Neoclass_Ecoc}
\bibitem{Neoclass_Ecoc}
\bibinfo{author}{E.~Roy \surnamestart Weintraub\surnameend}
  (\bibinfo{year}{2002}): \emph{\bibinfo{title}{Neoclassical Economics}}.
\newblock \bibinfo{publisher}{Library of Economics and Liberty}.
\newblock \bibinfo{note}{In David R. Henderson (ed.). \emph{Concise
  Encyclopedia of Economics} (1st ed.).}

\bibitemdeclare{misc}{wiki:centipede}
\bibitem{wiki:centipede}
\bibinfo{author}{\surnamestart Wikipedia\surnameend} (\bibinfo{year}{2015}):
  \emph{\bibinfo{title}{Centipede game --- Wikipedia{,} The Free
  Encyclopedia}}.
\newblock
  \urlprefix\url{http://en.wikipedia.org/w/index.php?title=Centipede_game&oldid=663077006}.
\newblock \bibinfo{note}{[Online; accessed 25-May-2015]}.

\end{thebibliography}

\end{document}